# Chaos-Based Bitwise Dynamical Pseudorandom Number Generator on FPGA

Miguel Garcia-Bosque, Adrián Pérez-Resa, Carlos Sánchez-Azqueta, Concepción Aldea, and Santiago Celma

*Abstract*— In this paper, a new pseudorandom number generator (PRNG) based on the logistic map has been proposed. To prevent the system to fall into short period orbits as well as increasing the randomness of the generated sequences, the proposed algorithm dynamically changes the parameters of the chaotic system. This PRNG has been implemented in a Virtex 7 field-programmable gate array (FPGA) with a 32-bit fixed point precision, using a total of 510 lookup tables (LUTs) and 120 registers. The sequences generated by the proposed algorithm have been subjected to the National Institute of Standards and Technology (NIST) randomness tests, passing all of them. By comparing the randomness with the sequences generated by a raw 32-bit logistic map, it is shown that, by using only an additional 16% of LUTs, the proposed PRNG obtains a much better performance in terms of randomness, increasing the NIST passing rate from 0.252 to 0.989. Finally, the proposed bitwise dynamical PRNG is compared with other chaos-based realizations previously proposed, showing great improvement in terms of resources and randomness.

*Index Terms*— Chaos, digital circuits, field-programmable gate array (FPGA), logistic map, pseudorandom number generator (PRNG), random number generation.

## I. Introduction

PSEUDORANDOM number generators (PRNG) have many applications among diverse fields such as cryptography [1], communications [2], or procedural generation [3]. Specifically, in the field of instrumentation and measurements, PRNGs are needed in many applications such as statistical sampling, Monte Carlo simulations, evaluating the immunity to noise of digital systems and, in general, testing of physical, biological, and electrical systems: code density tests and determination of Wiener and Volterra kernels in nonlinear systems [4], [5].

Some of the most commonly used PRNGs are based on linear congruential generators (LCG) or linear feedback shift registers (LFSR). Many of these systems, however, present some correlations or short periods, which make them unsuitable for many applications [6]. In this context, chaos-based

Manuscript received September 17, 2018; accepted October 12, 2018. This work was supported in part by MINECO-FEDER under Grant TEC2014-52840-R and Grant TEC2017-85867-R. The work of M. Garcia-Bosque was supported by FPU Fellowship under Grant FPU14/03523. The Associate Editor coordinating the review process was Leonid Belostotski. *(Corresponding author: Miguel Garcia-Bosque).*

The authors are with the Group of Electronic Design, Electrical Engineering and Communications Department, University of Zaragoza, 50009 Zaragoza, Spain (e-mail: mgbosque@unizar.es; aprz@unizar.es; csanaz@unizar.es; caldea@unizar.es; scelma@unizar.es).

Color versions of one or more of the figures in this paper are available online at http://ieeexplore.ieee.org.

Digital Object Identifier 10.1109/TIM.2018.2877859

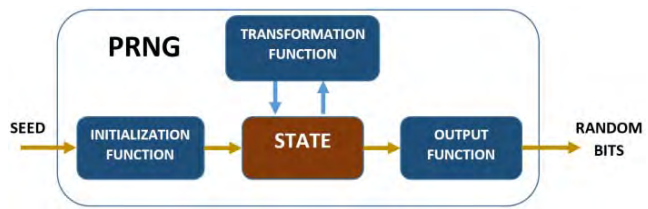

Fig. 1. Overall scheme of a generic PRNG.

PRNGs have arisen as a good alternative, thanks to their properties of ergodicity, and randomlike behavior [7].

In this paper, we propose a random generator based on the logistic map that, in order to improve its statistical properties, dynamically changes its chaotic parameter. The system has been implemented in a Virtex 7 field-programmable gate array (FPGA), using 510 lookup tables (LUTs) and 120 registers. To test the good statistical properties of the proposed generator, its generated sequences have been subjected to the National Institute of Standards and Technology (NIST) tests. The sequences have passed all of these tests, proving that they are undistinguishable from a truly random sequence.

The main contribution of this paper is the proposal of a novel chaos-based PRNG that:
1) offers better randomness results than other PRNGs commonly used in simulations such as LCGs and LFSRs;
2) requires a very small amount of resources to be implemented on an FPGA compared to other previously proposed chaos-based PRNGs.

## II. PRNG Algorithm

### A. Generic Structure of a Chaotic PRNG

A PRNG is an algorithm that, starting with a seed, by using a transformation function, generates a sequence that appears to be random, and its length is much bigger than the seed length (Fig. 1). A chaotic PRNG can be easily implemented by using a digitized chaotic map

$$x_{i+1} = f(x_i, \gamma) \tag{1}$$

where each $x_i$ is an element of the sequence and $\gamma$ is a constant parameter that determines the behavior of the system.

Using a map of this kind, starting from a seed composed by $x_0$ and $\gamma$, a sequence of elements $\{x_i\}$ is generated. Since each element is represented by a certain number of bits, it is possible to use all of them to build a binary random generator. However, there can be correlations among the bits within an element $x_i$. Therefore, to obtain better statistical properties, only a few bits of each $x_i$ are usually used to build the random sequences, typically the least significant bits (LSBs) since they present a low correlation.

In this paper, we have based the algorithm on the logistic map, given by

$$x_{i+1} = \gamma\, x_i(1 - x_i) \qquad (2)$$

where in order to work on the chaotic region, necessary for obtaining good random properties, the values of the parameter $\gamma$ must be in the interval [3.57, 4]. If $\gamma < 3.57$, fixed points or periodic orbits (not suitable for PRNG applications) are obtained and, if $\gamma > 4$, the orbits usually diverge [8]. This map has been chosen due to its simplicity and its randomlike behavior that has been widely studied in the past decades [9].

### B. Randomness Degradation Caused by the Digitization of the System

When a chaotic map such as the logistic map is digitized with a word length of $n$ bits, each $x_i$ can only take $2^n$ different values. Furthermore, for a given $\gamma$, the value of a certain $x_i$ determines the value of the next element $x_{i+1}$. Therefore, after a maximum number of $2^n$ iterations, the sequence will repeat itself. Although the maximum period is $2^n$, much shorter periods, on the order of $\sim 2^{n/2}$, are usually found of [10]. These short-period sequences fail most of the NIST randomness tests [8]. Furthermore, if a big number of random numbers were needed for signal processing or simulation, the sequence would start to repeat itself which could affect the simulation results.

A possible strategy to reduce this problem consists on using bigger word lengths. For example, a word length of 500 bits is used in [11]. Unfortunately, this approach requires to use a big amount of extra resources to obtain longer periods.

In this paper, this issue has been solved using an alternative approach that improves the random properties of a chaotic PRNG by using a small number of extra resources.

### C. Random-Enhancement Proposal

The random-enhancement approach used in this paper consists of using several values of $\gamma: \gamma_1, \gamma_2, \ldots \gamma_m$ instead of a single one. The sequence $\{x_i\}$ is generated by changing the value of $\gamma$ according to a sequence partition $\{k_i\}$. With this method, the first $k_1$ elements are obtained by $x_i = f(x_{i-1}, \gamma_1)$, the next $k_2$ elements are obtained as $x_i = f(x_{i-1}, \gamma_2)$, and so on. After having used all of the values of $\gamma_i$ and having generated $\sum_{i=1}^{m} k_i$ elements in total, the initial value of $\gamma$ and $\gamma_1$ is reused, continuing the process in a circular way. To prevent the system to fall into possible short cycles, the elements of $\{k_i\}$ are generated randomly within a certain range. Although this technique was originally advanced in [16] and [17] for a general case, in this paper, it has been improved by optimizing the values of $m$ and $\{k_i\}$ to obtain the best randomness improvement: while those works obtained an NIST passing rate up to 0.971, this paper obtain an NIST passing rate of 0.989. Furthermore, it has been applied to a simpler chaotic map than the skew tent map (used in [16] and [17]), the logistic map, obtaining a cost-effective high-performance PRNG.

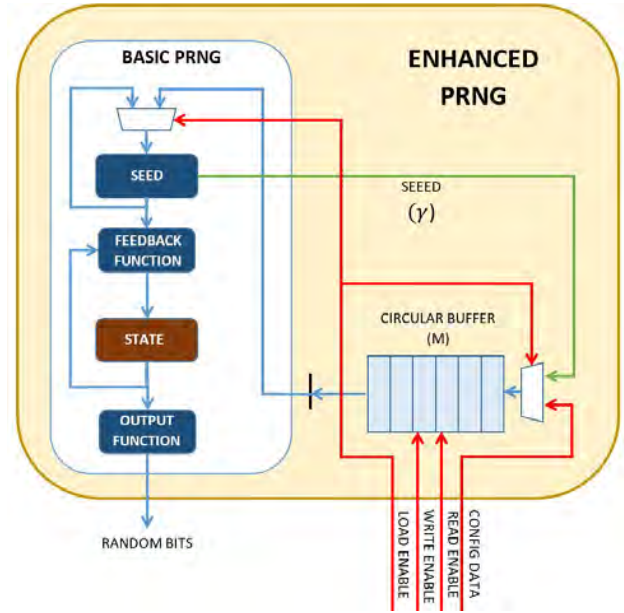

Fig. 2. Diagram of the enhanced PRNG. A FIFO is used to store the values of each $\gamma_i$. This is done with the config data signal. Write enable signal allows the system to write the value of config data inside the FIFO. Finally, read enable and load enable allow to read the next value of $\gamma_i$ from the FIFO and load it into the seed register. The "config. data," "read," "write," and "load" enable inputs are driven by an external control block (not shown in the picture) that has been designed to follow the proposed algorithm.

TABLE I
IMPLEMENTATION RESULTS

| Configuration | 32-bit Logistic Map | 64-bit Logistic Map | 32-bit Proposed PRNG |
|---|---|---|---|
| LUTs | 439 | 903 | 510 |
| Registers | 46 | 70 | 120 |
| Slices[1] | 116 | 235 | 143 |
| DSPs | 13 | 18 | 13 |
| NIST passing rate | 0.252 | 0.979 | 0.989 |

[1]The number of slices has been estimated from the number of registers and LUTs assuming that they are not packed together.

The proposed system has been fully implemented in an FPGA and exhaustively analyzed. A scheme of the enhanced PRNG is shown in Fig. 2.

### III. IMPLEMENTATION RESULTS

The proposed algorithm has been implemented in a Virtex 7 FPGA. A 32-bit word length has been used for the values of $\gamma_i$ and $x_i$ and a total of $m = 8$ different values of $\gamma_i$ have been used. The elements of the sequence partition have been obtained by generating random integers within the interval $k_i \in [9, 11]$ using a simple LCG algorithm. Finally, only the LSB of each $x_i$ has been used to generate the pseudorandom sequence.

To test the statistical properties of the PRNG, 100 sequences of $10^6$ bits have been generated and have been subjected to the NIST randomness tests [18], with a significance level of 0.01 (i.e., 99% of the sequences generated by a truly random generator would pass the tests). The NIST test results along with the implementation resources are given in Table I.

TABLE II
COMPARISON WITH OTHER CHAOTIC PRNGs

| System | [12] | [13] | [14] | [15] | Proposed PRNG |
|---|---|---|---|---|---|
| Platform | Virtex 6 | Virtex 6 | Virtex II | Spartan 3E | Virtex 7 |
| Main chaotic algorithm | Henon Map | Logistic Map | Lorenz System | Bernoulli Map | Logistic Map |
| LUTs | 1600 | 643 | 2718 | 575 | 510 |
| Registers | 64 | 160 | 791 | 108 | 120 |
| Slices[1] | 408 | 181 | 1755 | 342 | 143 |
| DSPs | 16 | 16 | 40[2] | 9 | 13 |
| NIST passing rate | NA | NA | NA | NA | 0.989 |
| Freq. (MHz) | 25.7 | 93 | 15.5 | 36.9 | 132 |
| Bits/cycle | 1 | 16 | 8 | 0.2 | 1 |

[1]Virtex II uses 18x18 multipliers instead of DSPs. [2]The number of slices has been estimated from the number of registers and LUTs assuming that they are not packed together.

To be able to compare the number of used resources with other previously proposed algorithms, the number of slices has been estimated from the number of LUTs and registers assuming unrelated logic. Since the NIST test suite provides a big number of tests, to summarize the results, only the average of the passing rates of all the tests is given in Table I (ideally, this value should be close to 0.99). For comparison purposes, the NIST tests results for the same logistic map with a 32 and a 64-bit precision using a single value of $\gamma$ are shown.

As it can be seen, while the passing rate of the 32-bit raw logistic map is very low, the NIST passing rate of the proposed system is very close to the ideal 0.99. While the 64-bit logistic map also obtains a high passing rate, it must be pointed out that the proposed system uses 39% fewer slices (that takes into account both LUTs and registers) as well as 28% fewer DSPs.

For comparison, the NIST test result for a glibc LCG (used by the GNU C Compiler) has a passing rate of 0.350 while a 32-order LFSR has a passing rate of 0.978, both of them lower than the proposed algorithm. Furthermore, these generators always fail some particular tests. In the case of the LFSR, all of the linear complexity test, as well as the random binary matrix tests, failed while, for the glibc LCG, all of the frequency, fast Fourier transform, cumulative sums, runs, overlapping template, approximate entropy, universal and serial failed. These can have a considerable effect in certain applications such as, for example, in Monte Carlo simulations, as proven in [6].

Finally, Table II shows a comparison among other previously proposed chaotic PRNGs. As it can be seen, the proposed PRNG shows great results in terms of resources, using the least amount of slices as well as the least amount of DSPs in the FPGA implementation. It must be pointed out that, although [13] and [14] achieve a higher throughput by transmitting 16 and 8 bits per cycle, respectively, they have not passed statistical tests as strict as the ones passed by the proposed PRNG (i.e., they have not proven that the passing rate of their generated sequences is close to 0.99).

## IV. CONCLUSION

In this paper, a new chaos-based bitwise dynamical PRNG has been proposed and tested. The system has proven to be capable of generating sequences with good statistical properties, passing the NIST randomness tests by using just a few more resources than the 32-bit logistic map generator. The proposed PRNG achieves better randomness than other commonly used PRNGs such as a 32-order LFSR or a glibc LCG.

Finally, a comparison of this PRNG with previously proposed chaos-based PRNGs proves the good performance of the proposed system, especially in terms of resources and quality of randomness.

This PRNG could be used in applications that do not require a high throughput but require a small area utilization or very good statistical properties such as, for example, Monte Carlo simulations.

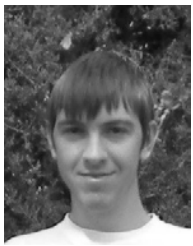
**Miguel Garcia-Bosque** was born in Zaragoza, Spain. He received the B.Sc. and M.Sc. degrees in physics from the University of Zaragoza, Zaragoza, in 2014 and 2015, respectively.

His current research interests include chaos theory and cryptography algorithms.

Dr. Garcia-Bosque is a member of the Group of Electronic Design, Aragón Institute of Engineering Research, University of Zaragoza.

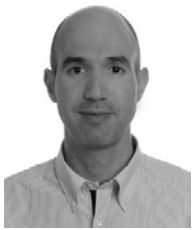
**Adrián Pérez-Resa** was born in San Sebastián, Spain. He received the M.Sc. degree in telecommunications engineering from the University of Zaragoza, Zaragoza, Spain, in 2005, where he is currently pursuing the Ph.D. degree with the Group of Electronic Design, Aragón Institute of Engineering Research.

He was a Research and Development Engineer at TELNET Redes Inteligentes, Zaragoza. His current research interests include high-speed communications and cryptography applications.

Dr. Pérez-Resa is a member of the Group of Electronic Design, Aragón Institute of Engineering Research, University of Zaragoza.

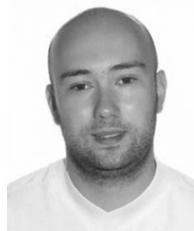
**Carlos Sánchez-Azqueta** was born in Zaragoza, Spain. He received the B.Sc., M.Sc., and Ph.D. degrees in physics from the University of Zaragoza, Zaragoza, Spain, in 2006, 2010, and 2012, respectively, and the Dipl.-Ing. degree in electronic engineering from the Complutense University of Madrid, Madrid, Spain, in 2009.

His current research interests include mixed-signal integrated circuits, high-frequency analog communications, and cryptography applications.

Dr. Sánchez-Azqueta is a member of the Group of Electronic Design, Aragón Institute of Engineering Research, University of Zaragoza.

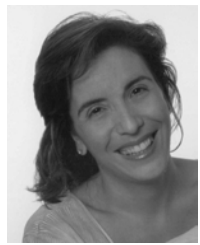
**Concepción Aldea** was born in Zaragoza, Spain. She received the B.Sc. and Ph.D. degrees in physics from the University of Zaragoza, Zaragoza, in 1990 and 2002, respectively.

She was with the private industry in the area of optical fiber research. She is currently an Associate Professor of electronics with the Faculty of Science, University of Zaragoza, where she is a Researcher with the Group of Electronic Design, Aragón Institute of Engineering Research. She is an Investigator of more than 30 national and international research projects. She has co-authored more than 20 technical papers and 70 international conference contributions. Her current research interests include mixed-signal IC design, continuous-time equalizers, and high-frequency optical fiber communication circuits.

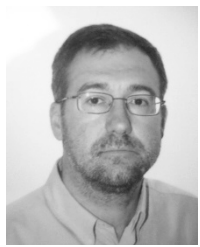
**Santiago Celma** was born in Zaragoza, Spain. He received the B.Sc., M.Sc., and Ph.D. degrees in physics from the University of Zaragoza, Zaragoza, in 1987, 1989, and 1993, respectively.

He is currently a Full Professor with the Group of Electronic Design, Aragon Institute of Engineering Research, University of Zaragoza. He is a Principal Investigator of more than 30 national and international research projects. He has co-authored more than 100 technical papers and 300 international conference contributions, four technical books. He holds four patents. His current research interests include circuit theory, mixed-signal integrated circuits, high-frequency communication circuits, wireless sensor networks, and cryptography for secure communications.